\documentclass[twocolumn,floatix,amssymb,amsmath,superscriptaddress,letterpaper]{revtex4}
\usepackage{hyperref}
\usepackage{graphicx}
\usepackage{color}
\usepackage{multibbl}

\begin{document}

\title{Simulating the exchange of Majorana zero modes with a photonic system}

\author{Jin-Shi Xu}

\affiliation{Key Laboratory of Quantum Information, University of Science and
Technology of China, CAS, Hefei, 230026, People's Republic of China}

\affiliation{Synergetic Innovation Center of Quantum Information and Quantum Physics,
University of Science and Technology of China, Hefei, Anhui 230026,
People's Republic of China}

\author{Kai Sun}

\affiliation{Key Laboratory of Quantum Information, University of Science and
Technology of China, CAS, Hefei, 230026, People's Republic of China}

\affiliation{Synergetic Innovation Center of Quantum Information and Quantum Physics,
University of Science and Technology of China, Hefei, Anhui 230026,
People's Republic of China}

\author{Yong-Jian Han}

\email{smhan@ustc.edu.cn}

\affiliation{Key Laboratory of Quantum Information, University of Science and
Technology of China, CAS, Hefei, 230026, People's Republic of China}

\affiliation{Synergetic Innovation Center of Quantum Information and Quantum Physics,
University of Science and Technology of China, Hefei, Anhui 230026,
People's Republic of China}

\author{Chuan-Feng Li}

\email{cfli@ustc.edu.cn}

\affiliation{Key Laboratory of Quantum Information, University of Science and
Technology of China, CAS, Hefei, 230026, People's Republic of China}

\affiliation{Synergetic Innovation Center of Quantum Information and Quantum Physics,
University of Science and Technology of China, Hefei, Anhui 230026,
People's Republic of China}

\author{Jiannis K. Pachos}

\affiliation{School of Physics and Astronomy, University of Leeds, Leeds LS2 9JT, United Kingdom}

\author{Guang-Can Guo}

\affiliation{Key Laboratory of Quantum Information, University of Science and
Technology of China, CAS, Hefei, 230026, People's Republic of China}

\affiliation{Synergetic Innovation Center of Quantum Information and Quantum Physics,
University of Science and Technology of China, Hefei, Anhui 230026,
People's Republic of China}

\date{\today }

\begin{abstract}
The realization of Majorana zero modes is in the centre of intense theoretical and experimental investigations. Unfortunately, their exchange that can reveal their exotic statistics needs manipulations that are still beyond our experimental capabilities. Here we take an alternative approach. Through the Jordan-Wigner transformation, the Kitaev's chain supporting two Majorana zero modes is mapped to the spin-1/2 chain. We experimentally simulated the spin system and its evolution with a photonic quantum simulator. This allows us to probe the geometric phase, which corresponds to the exchange of two Majorana zero modes positioned at the ends of a three-site chain. Finally, we demonstrate the immunity of quantum information encoded in the Majorana zero modes against local errors through the simulator. Our photonic simulator opens the way for the efficient realization and manipulation of Majorana zero modes in complex architectures.
\end{abstract}


\maketitle

Majorana zero modes (MZMs) give rise to a non-trivial Hilbert space encoded in their degenerate states. When two MZMs are exchanged, their degenerate subspace can be transformed by a unitary operator that signals their non-Abelian statistics~\cite{Wilczek1984}. In addition, quantum information encoded in the degenerate subspace of MZMs is naturally immune to local errors. The degeneracy of these states is protected by the fermion parity symmetry, so it cannot be lifted by any local symmetry-preserving perturbation. These unique characteristics make MZMs of interest to fundamental physics, while they can potentially provide novel and powerful methods for quantum information storing and processing~\cite{Nayak2008}.

The investigation of MZMs has attracted great attention. Rapid theoretical developments~\cite{Kitaev2000,Fu2008,Sau2010} have greatly simplified the technological requirements and made it possible to experimentally observe MZMs. A few positive signatures of the formation of MZMs have been reported recently in solid-state systems~\cite{Mourik2012,Deng2012,Rokhinson2012,Mebrahtu2013,Nadj-Perge2014,Lee2014}. Nevertheless, the central characteristic of exotic statistics requires the trapping and transport of MZMs in order to perform their exchange. This level of manipulation of MZMs is beyond our current control of solid-state systems.

In the following, we overcome this hurdle by taking advantage of the quantum simulation approach~\cite{Georgescu2013}: while the simulated system is not experimentally accessible with current technology, the quantum simulator and its measurement results provide information about the simulated system. We use a photonic quantum simulator to investigate the exchange of MZMs supported in the 1D Kitaev Chain Model (KCM). In principle, we can directly map the Fock space of the Majorana system to the space of the quantum simulator.  Nevertheless, to make our approach easily accessible to other scalable systems, such as trapped ions~\cite{Barreiro2011} or Josephson junctions~\cite{Devoret2013}, where spin modes are directly available, we divide the map into two consecutive steps. First, we perform the mapping of the Majorana system to the spin-1/2 system via the Jordan-Wigner (JW) transformation ~\cite{Kitaev2008,Paul2012,Alicea2015,Kells08}. Then we perform the mapping of the spin system to the spatial modes of single photons. The JW transformation~\cite{Jordan1928} is a non-local transformation that relates the states and evolutions of the topological superconducting Kitaev chain to the states and evolutions of a spin-1/2 chain. The corresponding spin system can then be simulated by our optical simulator. The simulated system has only two Majoranas (separated by a gap from all other excitations), any adiabatic manipulation can only result in a U(1)$\times$U(1) transformation that carries the hallmark of their anyonic statistics. In this way, we are able to demonstrate the exchange of two MZMs in a three-site Kitaev chain encoded in the spatial modes of photons. We further demonstrate that quantum information encoded in the degenerate ground state is immune to local phase and flip noise errors.

Due to the JW transformation, some of the non-local properties of the MZMs are lost in the simulated spin system. Indeed, local single-particle measurements can reveal the fusion space of the corresponding MZMs, something that is not possible in the fermionic picture.
At the same time, some of the local properties of the MZMs are also lost. For example, two isolated Majorana fermions, located at the end points of the chain, correspond to excitons in the spin system that are spread along the whole chain. Nevertheless, as the spectra of the two systems are the same, their corresponding quantum evolutions are equivalent. This gives us the means to investigate the statistical behaviour of the MZMs as well as their resilience to errors. Our current simulation only probes the Abelian part of the exchange as it employs two isolated MZMs. However, our technology and methodology can be used to demonstrate the non-Abelian statistics of Majorana fermions in which four MZMs are needed. This establishes our photonic quantum simulator as a novel platform to study the exotic physics of Majorana fermions and their possible applications to quantum information~\cite{Pachos2012}. Thus, our results extend the capabilities of optical simulators.

\section*{Results}
{\bf \noindent Majorana braiding.}
To clearly illustrate the properties of MZMs, we first consider the fermionic KCM~\cite{Kitaev2000}. This is the simplest model that supports two MZMs, $\gamma_{\rm A}$ and $\gamma_{\rm B}$ at its boundaries and exhibits two-fold degeneracy in its ground state. The braiding properties of Majorana fermions can be investigated by exchanging two MZMs. To perform such an exchange an adiabatic evolution between carefully constructed Hamiltonians ($H_{0}^{\rm MF},H_{1}^{\rm MF},\cdots,H_{n}^{\rm MF},H_{n+1}^{\rm MF}=H_{0}^{\rm MF}$) have been proposed to probe the statistics of the MZMs in the KCM~\cite{Alicea2011,Pachos2009}. The exchange of $\gamma_{\rm A}$ and $\gamma_{\rm B}$ induces the braiding evolution $U_\text{ex}=e^{{\it \pi}\gamma_{\rm B}\gamma_{\rm A}/4}$~\cite{Alicea2011} acting on the two-fold-degenerate ground-state space. As the MZMs belong to the same chain with a fixed fusion channel the braiding matrix is diagonal. Its elements are given in terms of geometric phases~\cite{Berry1984} on the basis of $|0_{\rm Lf}\rangle$ (odd fermion parity) and $|1_{\rm Lf}\rangle$ (even fermion parity), where the subscript $\rm f$ denotes fermionic states and $\rm L$ is the number of sites of the chain (see section IA in the Supplementary Information (SI) for more details). According to the work of Pancharatnam~\cite{Pancharatnam1956} and the Bargmann invariants~\cite{Bargmann1964}, the geometric phases resulting from the braiding process can be directly determined through projective measurements
\begin{equation}
\varphi_\text{g}=-\text{arg}(\langle m_{\rm Lf}|P_{1}P_{2}\cdots P_{n}|m_{\rm Lf}\rangle),
\label{eq:geometricphase}
\end{equation}
where $\varphi_\text{g}$ is the geometric phase associated with $|m_{\rm Lf}\rangle$ which represent the basis for the ground states of the Hamiltonian $H_{0}^{\rm MF}$ ($m=0$ or $1$). Here, $P_{j}$ projects the system into the ground
space of $H_{j}^{\rm MF}$, for $j=1,...,{\rm n}$. The geometric phases are gauge invariant and
uniquely determined by the Hamiltonians $H_{0}^{\rm MF},H_{1}^{\rm MF},\cdots,H_{n}^{\rm MF}$,
while the details of the adiabatic processes between them are not essential~\cite{Simon1993,Aharonov1987}.
Generally, the projective measurement can be expressed as an imaginary-time
evolution (ITE) operator $e^{-H_{j}^{\rm MF}t}$ with a sufficiently large
evolution time $t$ ~\cite{Vidal2007}. The adiabatic requirement preserves the fermion parity
of the initial state and guarantees that the off-diagonal elements
in $U_\text{ex}$ remain zero~\cite{Simon1993}. Therefore,
the whole information of $U_\text{ex}$ can be read out from ITE operators
$e^{-H_{1}^{\rm MF}t}e^{-H_{2}^{\rm MF}t}\cdots e^{-H_{n}^{\rm MF}t}$~\cite{kapit2012}.
By employing the method of dissipative evolution~\cite{Verstraete2009},
each projective measurement can be efficiently completed with some
finite probability (a general circuit is given in section IB in SI). We shall employ this method to experimentally probe the geometric phases obtained during the exchange of two MZMs by simulating a spin model that is equivalent to the KCM.

\begin{figure*}[tbph]
\begin{centering}
\includegraphics[width=1\columnwidth]{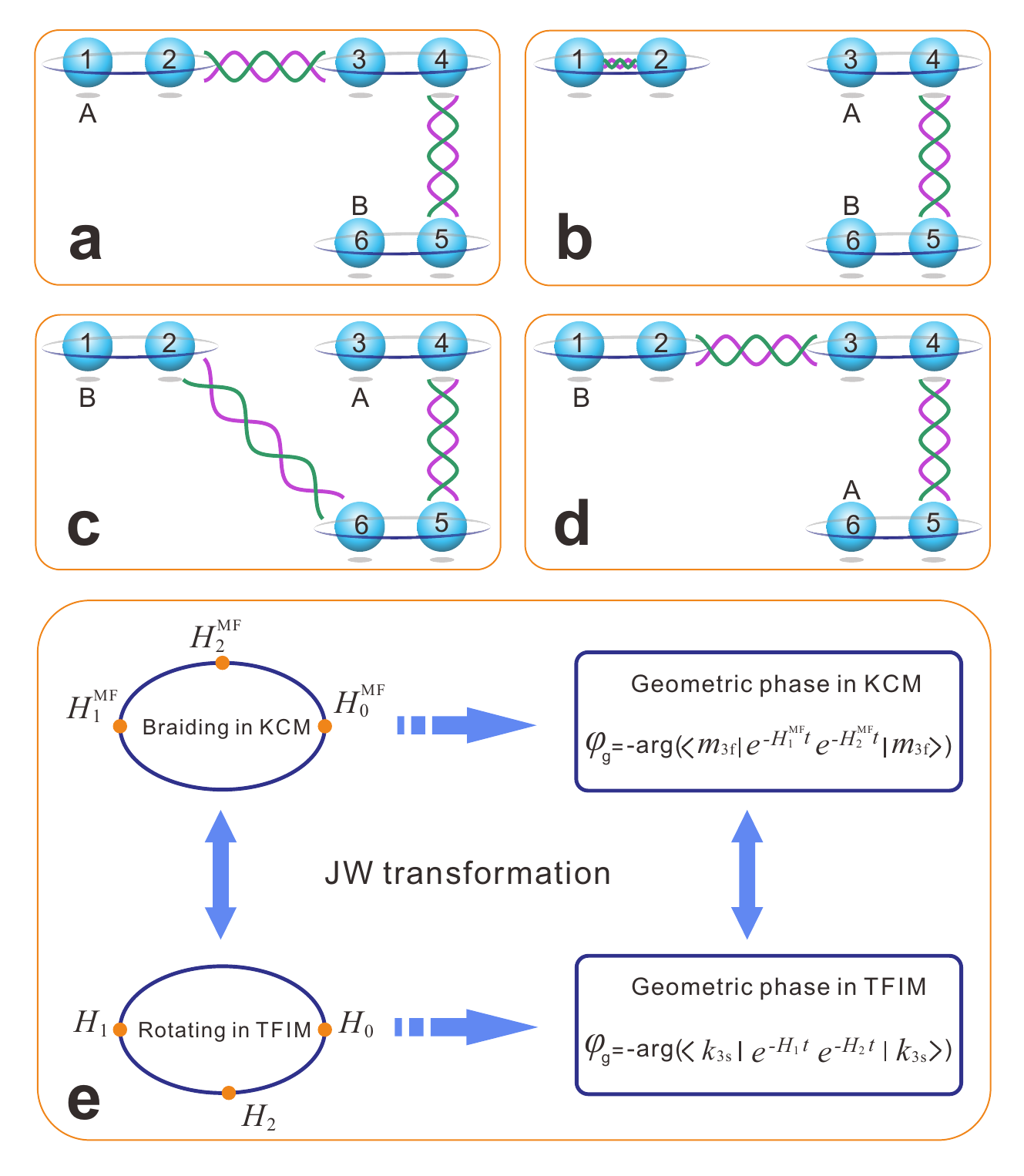}
\par\end{centering}

\protect\caption{The braiding of Majorana zero modes and the mapping between the fermionic and spin models. The spheres with numbers at their centres represent the Majorana fermions, $\gamma_j$, for $j=1,...,6$, at the corresponding sites. A pair of Majorana
fermions bounded by an enclosing ring represents a normal fermion.
The wavy lines between different sites represent the interactions
between them. The interactions illustrated in \textbf{a}, \textbf{b},
\textbf{c} and \textbf{d} represent the Hamiltonians $H_{0}^{\rm MF}$,
$H_{1}^{\rm MF}$, $H_{2}^{\rm MF}$ and $H_{0}^{\rm MF}$, respectively. The
letters $\rm A$ and $\rm B$ in each pane represent the corresponding isolated
Majorana zero modes. The mapping between the Kitaev chain model (KCM) and
the transverse-field
Ising model (TFIM) through the Jordan-Wigner (JW) transformation is shown in \textbf{e}. $m_{\rm 3f}$ and $k_{\rm 3s}$ are bases of the ground-state spaces defined
in the KCM and TFIM ($m$ and $k$=0 or 1), respectively.}
\label{fig:map}
\end{figure*}

To complete the exchanging of two MZMs, a four-fermion scheme has
been proposed in the superconducting wire system~\cite{Alicea2011},
which is further reduced to a three-fermion scheme in our work. The
three-fermion KCM can be described in terms of six Majorana fermions,
denoted by $\gamma_{j}$ ($\gamma^\dag_{j}=\gamma_{j}$, $\gamma_{k}\gamma_{l}+\gamma_{l}\gamma_{k}=2\delta_{k,l}$ for $j,k,l=1,...,6$), as shown in Figure ~\ref{fig:map}. The exchanging process can be completed
by a set of adiabatic processes among three different Hamiltonians,
as illustrated in Figure~\ref{fig:map}\textbf{a-d}. The initial
Hamiltonian is $H_{0}^{\rm MF}=i(\gamma_{2}\gamma_{3}+\gamma_{4}\gamma_{5})$.
The MZMs $\gamma_{1}$ and $\gamma_{6}$ are isolated at the boundary of the
chain to form two MZMs, namely $\rm A$ and $\rm B$ (see Figure~\ref{fig:map}\textbf{a}),
and the degenerate ground-state space of $H_{0}^{\rm MF}$ is spanned
by $|0_{\rm 3f}\rangle$ and $|1_{\rm 3f}\rangle$
($\rm L=3$). The other two Hamiltonians are as follows: $H_{1}^{\rm MF}=i(\gamma_{4}\gamma_{5}+\frac{1}{2}\gamma_{1}\gamma_{2})$,
where $\gamma_{3}$ and $\gamma_{6}$ are isolated to form MZMs (see
Figure~\ref{fig:map}\textbf{b}), and $H_{2}^{\rm MF}=i(\gamma_{4}\gamma_{5}+\gamma_{2}\gamma_{6})$,
where $\gamma_{1}$ and $\gamma_{3}$ are isolated to form MZMs (see
Figure~\ref{fig:map}\textbf{c}). The MZMs cross during the adiabatic
transition from $H_{1}^{\rm MF}$ to $H_{2}^{\rm MF}$. To complete the exchange
process, we adiabatically transform $H_{2}^{\rm MF}$ back into $H_{0}^{\rm MF}$,
and the Majorana mode $\rm A$ is driven to the location of $\gamma_{6}$
(see Figure~\ref{fig:map}\textbf{d}). The exchange of $\rm A$ and $\rm B$
is thus completed, introducing the unitary operation $U_\text{ex}=e^{\pi\gamma_{1}\gamma_{6}/4}$
in the ground-state space on the basis of $|0_{\rm 3f}\rangle$ and $|1_{\rm 3f}\rangle$.

Through the JW transformation, a general KCM can be transformed into a 1D transverse-field Ising model (TFIM)~\cite{Kitaev2008}. However, these two models have some differences in their physics (see section IC in SI for more details). In the fermionic system, the total fermion parity is fixed. The braiding effect in a single KCM is an overall phase. This phase cannot be directly measured as the superposition state with different fermionic parity is impossible. Despite that, the state of the spin model can be prepared in any superposition state. Hence, the relative geometric phase obtained during the exchange can be measured. As the KCM and the TFIM in the ferromagnetic region have the same spectra and their corresponding quantum evolutions are equivalent, the geometric phases obtained from the braiding evolution $U_{\rm{ex}}$ are invariant under this mapping (see section IE in SI for more details). As a result, the well-controlled spin system offers a good platform to determine the exchange matrix and investigate the statistical behaviour of MZMs.

The Hamiltonians involved in the three-fermion braiding scheme of the KCM, i.e., $H_{0}^{\rm MF}$, $H_{1}^{\rm MF}$ and $H_{2}^{\rm MF}$, are
transformed through the JW transformation into a spin-1/2 system as follows (see section ID in SI for more details):
\begin{equation}
\begin{split}H_{0} & =-(\sigma_{1}^{x}\sigma_{2}^{x}+\sigma_{2}^{x}\sigma_{3}^{x})\quad,\\
H_{1} & =-\sigma_{2}^{x}\sigma_{3}^{x}+\frac{1}{2}(\sigma_{1}^{z}+1)\quad,\\
H_{2} & =-(\sigma_{2}^{x}\sigma_{3}^{x}+\sigma_{1}^{x}\sigma_{2}^{z}\sigma_{3}^{y})\quad,
\end{split}
\label{eq:Hamiltonian}
\end{equation}
where $\sigma_{i}^{x}$, $\sigma_{i}^{y}$ and $\sigma_{i}^{z}$ are Pauli matrices of the $i$th spin. The ground states of these Hamiltonians are two-fold degenerated. The basis of the ground states of $H_{0}$ are denoted by $|0_{\rm 3s}\rangle$
and $|1_{\rm 3s}\rangle$, ($\rm s$ indicates spin states) corresponding to the basis of $|0_{\rm 3f}\rangle$
and $|1_{\rm 3f}\rangle$ in KCM through JW transformation, respectively.
The detailed form of $U_\text{ex}=e^{-H_{2}t}e^{-H_{1}t}$ in the ground space
of $H_{0}$, which is the same as $e^{-H_{2}^{\rm MF}t}e^{-H_{1}^{\rm MF}t}$
in the ground space of $H_{0}^{\rm MF}$, can be
obtained by the tomography process, where
the initial ground states of $H_{0}$ are obtained from a randomly
chosen input state $|\phi_{\rm r}\rangle$ by implementing $e^{-H_{0}t}$
with a sufficiently large $t$, i.e., $e^{-H_{0}t}|\phi_{\rm r}\rangle$
(not normalized). The relative geometric phases and the exchange
property of MZMs in the corresponding KCM can then be deduced from
the information of $U_\text{ex}$. The mapping between the three-particle
KCM and TFIM is further shown in Figure~\ref{fig:map}\textbf{e}.

{\bf \noindent Local noise immunity.}
The robustness of the information encoded in the ground space of the KCM with Hamiltonian $H_{0}^{\rm MF}$ can also be studied in our system. We note that the degeneracy of the ground states of $H_{0}$ is not topologically protected, as it can be lifted by a local operator that breaks $\rm Z_{2}$ symmetry, while the ground state of KCM is topologically protected~\cite{Greiter2014}. However, due to the equivalence of the spectra of the $H_{0}^{\rm MF}$ and $H_{0}$, the information encoded in the ground state of $H_{0}$ is protected by the same excitation gap against local noises with $\rm Z_{2}$ symmetry which correspond to the noises in fermionic system.
It is worth noting that, in our work, the ground states obtained from the imaginary-time evolution process are algorithmically encoded like in quantum error correction~\cite{Freeman2016}. This non-local encoding of states in a larger number of physical qubits protects them against local flip and phase errors. In other words, the Ising model is a good memory with some of the robustness characteristics of the Majorana~\cite{Freeman2016}. As a result, the immunity of the local noises in KCM can be faithfully investigated, which establishes it as an attractive medium for storing and processing quantum information \cite{Kitaev2008}.

Without loss of generality, we take the local error $D_{j}$ acting on a certain
site $j$ of the KCM. To verify the robustness against this local
error, we should confirm that the operator $D_{j}$ has a trivial
effect on any state in the ground-state space of $H_{0}^{\rm MF}$, i.e.,
$\langle\phi_{\rm f}|{D}_{j}|\phi_{\rm f}\rangle$ is constant for all the states
$|\phi_{\rm f}\rangle$ in the ground-state space of $H_{0}^{\rm MF}$. There are two independent types of local noise: flip error (${X}_{j}$) and phase error (${Z}_{j}$).
According to the argument presented in~\cite{Kitaev2000}, the single local
flip error is impossible due to the fermionic parity conversation
in the ideal KCM (we do not consider the quasiparticle poisoning effect in practical systems~\cite{Rainis2012}).
Therefore, the local flip error can only happen
at two sites simultaneously, such as $X_{1}=c_{1}^{\dag}c_{2}$ ($c_{j}$
and $c_{j}^{\dag}$ are the annihilation and creation operators of
the $j$th fermions, respectively). In terms of the TFIM the $X_{1}$ error operator becomes $\frac{1}{4}(i\sigma_{1}^{y}\sigma_{2}^{x}+\sigma_{1}^{y}\sigma_{2}^{y}+\sigma_{1}^{x}\sigma_{2}^{x}-i\sigma_{1}^{x}\sigma_{2}^{y})$
through the JW transformation. Meanwhile, the local
phase error on site 1, $Z_{1}=c_{1}^{\dag}c_{1}$, corresponds to
the operator $\frac{1}{2}(\sigma_{1}^{z}+1)$ of the spin system.
Hence, to confirm the robustness of local errors in KCM, we need to check
that $\langle\phi_{\rm 3f}|c_{1}^{\dag}c_{2}|\phi_{\rm 3f}\rangle=\frac{1}{4}\langle\phi_{\rm 3s}|(i\sigma_{1}^{y}\sigma_{2}^{x}+\sigma_{1}^{y}\sigma_{2}^{y}+\sigma_{1}^{x}\sigma_{2}^{x}-i\sigma_{1}^{x}\sigma_{2}^{y})|\phi_{\rm 3s}\rangle$
and $\langle\phi_{\rm 3f}|c_{1}^{\dag}c_{1}|\phi_{\rm 3f}\rangle=\langle\phi_{\rm 3s}|\frac{1}{2}(\sigma_{1}^{z}+1)|\phi_{\rm 3s}\rangle$ are constant
for all states $|\phi_{\rm 3s}\rangle$ in the ground-state space of $H_{0}$ (corresponding to all states $|\phi_{\rm 3f}\rangle$ in the ground state space of $H_{0}^{\rm MF}$).

{\bf \noindent Experimental results on simulating the braiding evolution.}
In our experiment the ITE operator plays a central role. For our current purpose, a useful method to implement
the ITE is to design appropriate dissipative evolution~\cite{Verstraete2009},
in which the ground-state information of the corresponding Hamiltonian
is preserved but the information of the other states is dissipated (see Methods).
Taking advantage of the commutation between $\sigma_{1}^{x}\sigma_{2}^{x}$ and
$\sigma_{2}^{x}\sigma_{3}^{x}$, $\sigma_{2}^{x}\sigma_{3}^{x}$ and
$\frac{1}{2}(\sigma_{1}^{z}+1)$, and $\sigma_{2}^{x}\sigma_{3}^{x}$
and $\sigma_{1}^{x}\sigma_{2}^{z}\sigma_{3}^{y}$ in Equation (\ref{eq:Hamiltonian}),
the operator $e^{-H_{2}t}e^{-H_{1}t}e^{-H_{0}t}$ can be expressed as
\begin{equation}
\begin{split}
 &e^{-H_{2}t}e^{-H_{1}t}e^{-H_{0}t} \quad \\ &=e^{(\sigma_{2}^{x}\sigma_{3}^{x}+\sigma_{1}^{x}\sigma_{2}^{z}\sigma_{3}^{y})t}e^{(\sigma_{2}^{x}\sigma_{3}^{x}-\frac{1}{2}(\sigma_{1}^{z}+1))t}e^{(\sigma_{2}^{x}\sigma_{3}^{x}+\sigma_{1}^{x}\sigma_{2}^{x})t}\quad\\
 & =e^{\sigma_{2}^{x}\sigma_{3}^{x}t}e^{\sigma_{1}^{x}\sigma_{2}^{z}\sigma_{3}^{y}t}e^{\sigma_{2}^{x}\sigma_{3}^{x}t}e^{-\frac{1}{2}(\sigma_{1}^{z}+1)t}e^{\sigma_{2}^{x}\sigma_{3}^{x}t}e^{\sigma_{1}^{x}\sigma_{2}^{x}t},
\end{split}
\label{eq:commutation}
\end{equation}
where we set $t=5$ in our analysis.
Due to the fact that the eigenvectors and eigenvalues of each local
operator, such as $\sigma_{2}^{x}\sigma_{3}^{x}$, can be obtained
easily, the operation of $e^{\sigma_{2}^{x}\sigma_{3}^{x}t}$ can
be implemented by the basis rotation (local unitary gate operation) and dissipation straightforwardly (see section IF in SI for more details). We therefore introduce an environmental
degree of freedom and couple it to the system, which can be simply
written as $|\phi_{\rm s}\rangle=\frac{1}{\sqrt{2}}(|\phi_{\rm g}\rangle|0_{\rm e}\rangle+|\phi_{\rm g}^{\perp}\rangle|1_{\rm e}\rangle)$, where
$|\phi_{\rm g}^{\perp}\rangle$ denotes the states that are orthogonal
to the ground state $|\phi_{\rm g}\rangle$. The environmental basis of
$|1_{\rm e}\rangle$ is dissipated during the evolution, and only $|0_{\rm e}\rangle$
is preserved. Therefore, the ground state of the corresponding Hamiltonian
is obtained in the digital quantum simulator~\cite{Barz2015}. It is worth to emphasize that this process is only dependent
on the Hamiltonian. The general algorithm of ITE for a general
Hamiltonian has been demonstrated in investigating algorithmic quantum
cooling~\cite{Xu2014}.

Here we employ ITE in optical systems, that provide ideal platforms for quantum simulations. They offer good control of their external parameters
and they are largely isolated by their environment~\cite{Alan2012}. As a proof-of-principle demonstration we map the three-fermion system to the spin model and
encode their information in spatial modes. In this way we can
experimentally probe the braiding matrix
in an all-optical system.

\begin{figure*}[tbph]
\begin{centering}
\includegraphics[width=2\columnwidth]{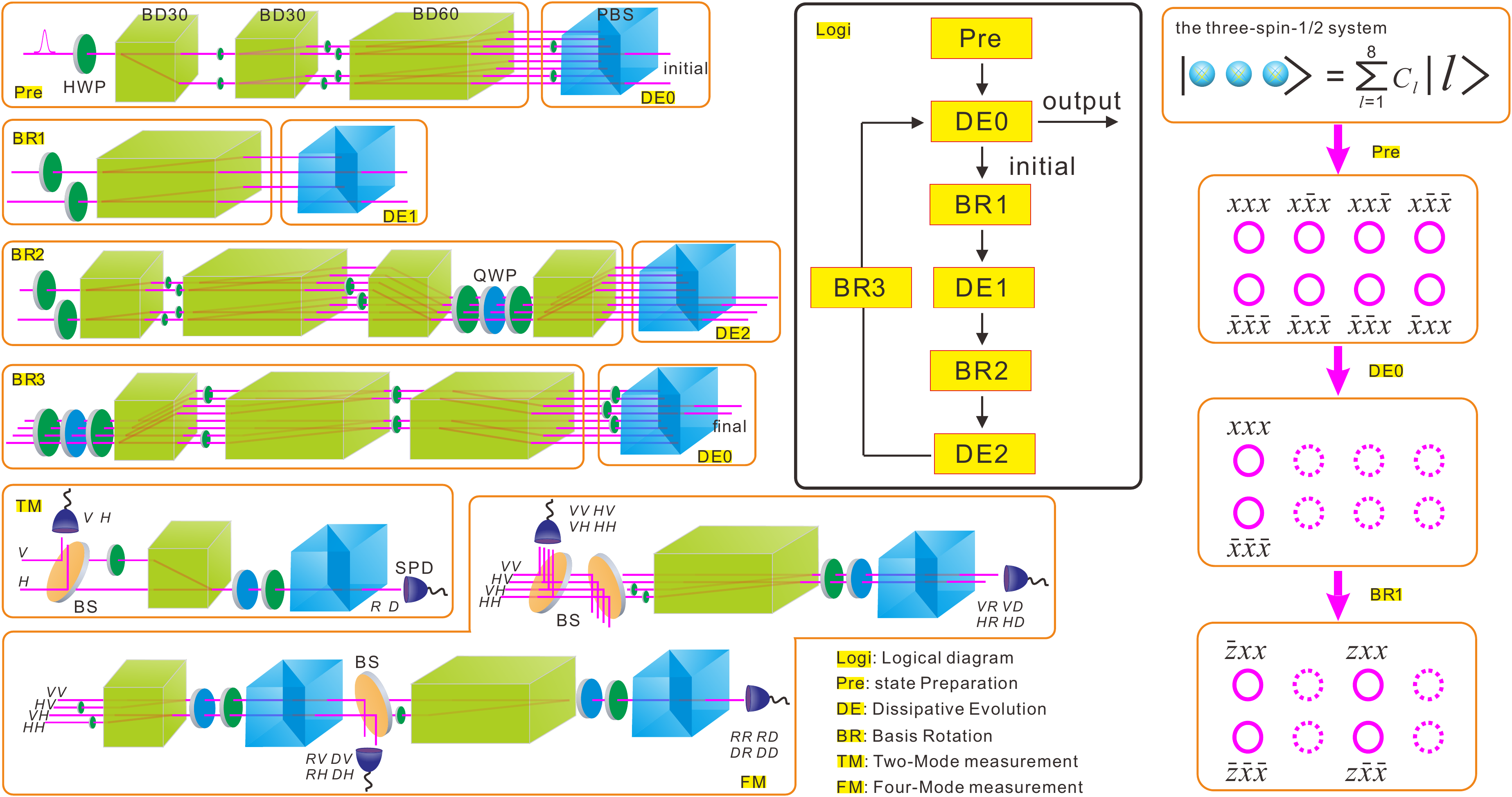}
\par\end{centering}

\protect\caption{Experimental setup for the simulation of the exchange of Majorana zero modes. The process follows the logical diagram provided in the pane
enclosed by the black solid line, denoted by Logi. The state of the three-spin-1/2 system can be expressed in the eight dimensional space with the basis denoted by $|l\rangle$ ($l=1$ to $8$), which are encoded as the spatial modes of photons. $C_{l}$ are the corresponding amplitudes. For the initial Hamiltonian $H_{0}$, the space is expanded by the basis of $|xxx\rangle, |x\bar{x}x\rangle, |xx\bar{x}\rangle, |x\bar{x}\bar{x}\rangle, |\bar{x}\bar{x}\bar{x}\rangle, |\bar{x}x\bar{x}\rangle, |\bar{x}\bar{x}x\rangle, |\bar{x}xx\rangle$. The polarization
of single photons is rotated using half-wave plates (HWPs) and quarter-wave plate (QWPs), and the
spatial modes are separated by beam displacers, each with a beam displacement
of either 3.0 mm (BD30) or 6.0 mm (BD60). The state preparation is
illustrated in the pane labeled Pre. The basis rotations BR1, BR2
and BR3 are used to express the input states in terms of the eigenstates
of $H_{1}$, $H_{2}$ and $H_{0}$, respectively. The dissipative
evolutions DE0, DE1 and DE2 drive the input states to the ground states
of $H_{0}$, $H_{1}$ and $H_{2}$, respectively. Some of the detailed basis representations of the spatial modes are given in the right column. The solid magenta rings represent the preserved optical modes, and the dashed magenta rings represent the discarded optical modes. The states indicated near the optical modes represent the corresponding preserved basis in the eight-dimensional space. Two types of measurements
are performed, i.e., two-mode measurement (TM) and four-mode measurement
(FM). Beam splitters (BSs) are used to send the photons to different
measurement instruments. $H$, $V$, $R$ and $D$ represent the horizontal polarization, vertical polarization, right-hand circular polarization and diagonal polarization, respectively. Finally, photons are detected using single-photon
detectors (SPDs).}
\label{fig:setup_nonabelian}
\end{figure*}

Figure~\ref{fig:setup_nonabelian} shows the experimental setup for
the investigation of the MZMs (see Methods). Our quantum simulator is realized by single photons. The states of three spin-$1/2$ sites correspond to a $2^{3}$-dimensional Hilbert space, which are encoded
in the optical spatial modes of single photons. The spatial propagation of the photon in one direction is equivalent to the imaginary time-evolution of (lower-dimensional) system.
It is worth to note that during the mapping between the many-body-to-single-particle Hilbert spaces the notion of locality is lost: excitations that are nearby in the system that is simulated, do not correspond to nearby spatial modes in the photonic system. Nevertheless, there are intriguing similarities.
The basis of the spatial modes of photons is the same as that of the corresponding Hilbert space of three-spin system (see sections IF and IG in SI for more details).

The basis of a spin state can be expressed as the eigenstates of $\sigma^{z}$ (denoted by $|z\rangle$,
with an eigenvalue of 1, and $|\bar{z}\rangle$, with an eigenvalue of $-1$),
$\sigma^{y}$ (denoted by $|y\rangle$, with an eigenvalue of 1, and
$|\bar{y}\rangle$, with an eigenvalue of $-1$) or $\sigma^{x}$ (denoted
by $|x\rangle$, with an eigenvalue of 1, and $|\bar{x}\rangle$, with an
eigenvalue of $-1$). In our experiment, we employ beam displacers
(BDs) of various lengths to prepare the initial states. A beam displacer
is a birefringent crystal, in which light beams with horizontal and
vertical polarizations are separated by a certain displacement (depending
on the length of the crystal)~\cite{Kitagawa2012}. Two types of
BDs are employed, BD30 (with a beam displacement of 3.0 mm)
and BD60 (with a beam displacement of 6.0 mm). The polarization of
the photon can be rotated using half-wave plates (HWPs), and the relative
amplitudes of the different spatial modes can be conveniently adjusted.
To obtain the ground states of the corresponding Hamiltonian, the
polarization of the photons is used as the environmental degree of
freedom. The coupling between the spatial mode and the polarization
is achieved using HWPs, which rotate the polarization in the corresponding
paths. Dissipative evolution is accomplished by passing the photons
through a polarization beam splitter (PBS), which transmits the horizontal
component and reflects the vertical component. In our case, only photons
in the optical modes that have horizontal polarizations are preserved;
these modes correspond to the ground states of the Hamiltonian. The
components with vertical polarizations are completely dissipated. The precision of the dissipative evolution is dependent on the ratio between the reflected and transmitted parts of the vertical polarization after the PBS, which can be higher than 500:1.
To ensure dissipative dynamics using a PBS, we must express the input
states in the basis of the eigenstates of the corresponding Hamiltonian,
which are referred to basis rotations. The eigenstates of $H_{0} =-(\sigma_{1}^{x}\sigma_{2}^{x}+\sigma_{2}^{x}\sigma_{3}^{x})$ can be expressed as $\{|xxx\rangle, |xx\bar{x}\rangle, |x\bar{x}x\rangle, |x\bar{x}\bar{x}\rangle, |\bar{x}xx\rangle, |\bar{x}x\bar{x}\rangle, |\bar{x}\bar{x}x\rangle, |\bar{x}\bar{x}\bar{x}\rangle\}$, which are represented by the eight spatial modes of the single photon and are shown in the state preparation pane labelled by Pre in Figure~\ref{fig:setup_nonabelian}. The corresponding cross sections of the spatial modes are shown in the right column in Figure~\ref{fig:setup_nonabelian}. Only the terms of $|xxx\rangle$ and $|\bar{x}\bar{x}\bar{x}\rangle$ are preserved after the ITE operation of $H_{0}$, which corresponds to two spatial modes in the dissipative evolution of DE0. For the ITE operation of $H_{1} =-\sigma_{1}^{x}\sigma_{2}^{x}+1/2(\sigma_{1}^{z}+1)$, we should only consider the term of $1/2(\sigma_{1}^{z}+1)$ (the term of $-\sigma_{1}^{x}\sigma_{2}^{x}$ is the same as that in $H_{0}$). Two HWPs with the angles setting to be $45^{\circ}$ in the basis rotation BR1 are implemented in modes of $|xxx\rangle$ and $|\bar{x}\bar{x}\bar{x}\rangle$, in which the basis of the first spin is rotated to $|z\rangle$ and $|\bar{z}\rangle$. The two spatial modes are then separated to four by a BD60 with the spatial modes represented as $|\bar{z}xx\rangle$, $|zxx\rangle$, $|\bar{z}\bar{x}\bar{x}\rangle$ and $|z\bar{x}\bar{x}\rangle$. After the dissipative evolution of DE1, only the terms of $|\bar{z}xx\rangle$ and $|\bar{z}\bar{x}\bar{x}\rangle$ are preserved. The ITE operation of $H_{2}$ can be implemented by the same way with the details given in the section IF in SI..

\begin{figure*}[tbph]
\begin{centering}
\includegraphics[width=1\columnwidth]{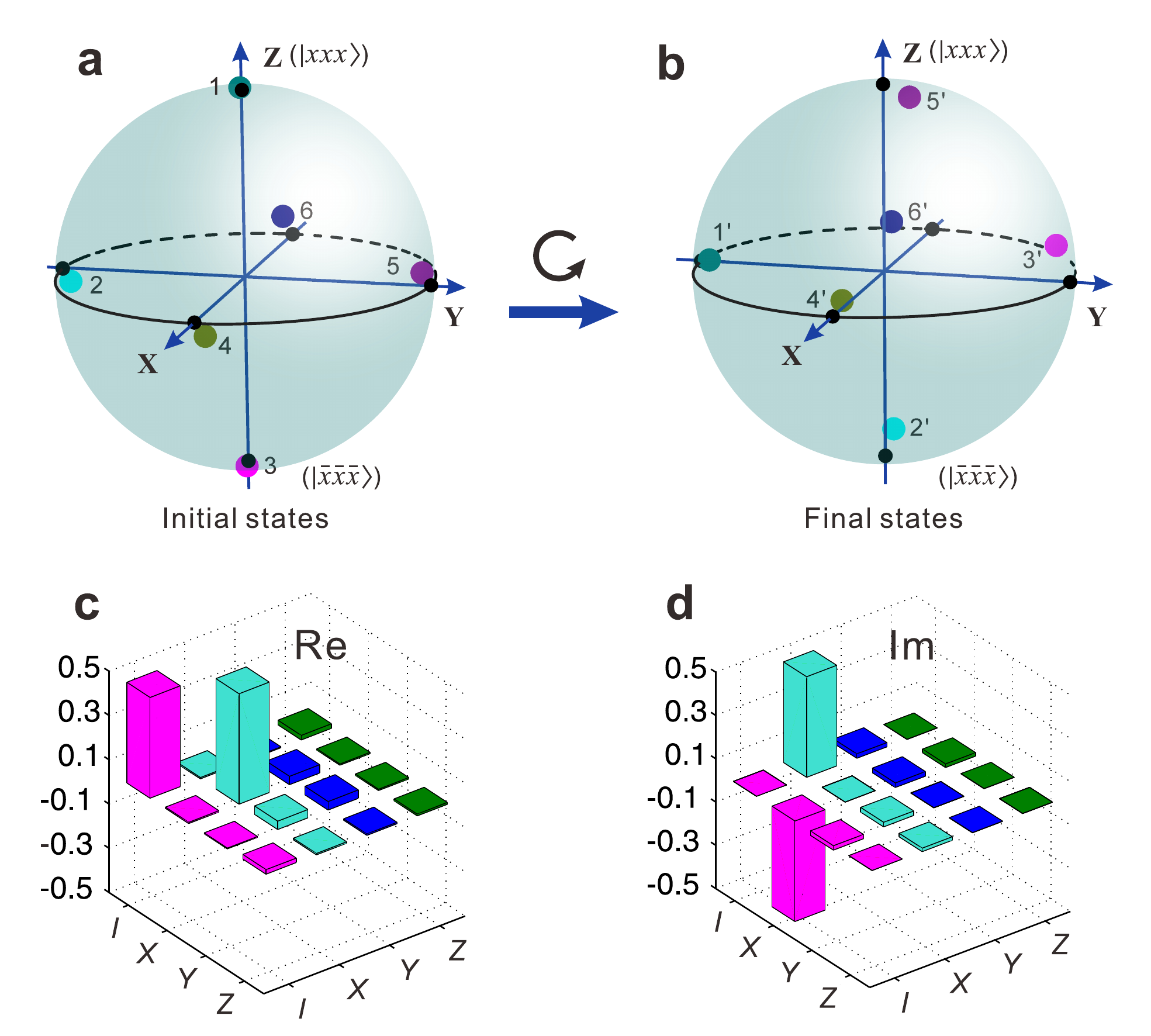}
\par\end{centering}

\protect\caption{Experimental results on simulating the braiding evolution. \textbf{a.} The six experimental initial states after the first dissipative evolution DE0 with the dark green dot labeled as 1, cyan dot labeled as 2, magenta dot labeled as 3, dark yellow dot labeled as 4, violet dot labeled as 5 and navy dot labeled as 6 in the Bloch sphere. \textbf{b.} The corresponding experimental final states after the second DE0 with the dark green dot labeled as $1'$,
cyan dot labeled as $2'$, magenta dot labeled as $3'$,
dark yellow dot labeled as $4'$, violet dot labeled as
$5'$, and navy dot labeled as $6'$ in the Bloch sphere. The black dots in the poles of the Bloch spheres represent the corresponding theoretical predictions with the states $|xxx\rangle$ (+{\bf Z} direction), $\frac{1}{\sqrt{2}}(|xxx\rangle-i|\bar{x}\bar{x}\bar{x}\rangle)$ (-{\bf Y} direction),
$|\bar{x}\bar{x}\bar{x}\rangle$ (-{\bf Z} direction), $\frac{1}{\sqrt{2}}(|xxx\rangle+|\bar{x}\bar{x}\bar{x}\rangle)$ (+{\bf X} direction, $|0_{\rm 3s}\rangle$),
$\frac{1}{\sqrt{2}}(|xxx\rangle+i|\bar{x}\bar{x}\bar{x}\rangle)$ (+{\bf Y} direction), and $\frac{1}{\sqrt{2}}(|xxx\rangle-|\bar{x}\bar{x}\bar{x}\rangle)$
(-{\bf X} direction, $|1_{\rm 3s}\rangle$), respectively. Due to the experimental errors, the coloured dots (experimental results) are slightly separated from the corresponding black dots. The final states are shown to be rotated along the {\bf X} axis by $\pi/2$ from the initial states. \textbf{c.} Real (Re) and \textbf{d.} Imaginary (Im)
parts of the exchange operator in the basis of $\{{I} ({\rm identity}), {X} (\sigma^{x}), {Y} (\sigma^{y}), {Z} (\sigma^{z})\}$, with a fidelity of $94.13\pm0.04$\%.}
\label{fig:finalstate}
\end{figure*}

To clearly show the two-mode output states, the corresponding density
matrices, $\rho$, are expressed in the basis of $\{I,\sigma^{x},\sigma^{y},\sigma^{z}\}$
as follows: $\rho=\frac{1}{2}(I+p_{1}\sigma^{x}+p_{2}\sigma^{y}+p_{3}\sigma^{z})$. Here $I$ represents the identity operator and $p_{1}$, $p_{2}$ and $p_{3}$
are the corresponding real amplitudes, which uniquely identify density
matrices on a Bloch sphere. The initial states after the ITE operation of DE0 is shown in Figure~\ref{fig:finalstate}\textbf{a}. The corresponding final states after the operator $U_\text{ex}$
are illustrated in Figure~\ref{fig:finalstate}\textbf{b}. It is shown that the final state is the same
as the initial state when the initial state is $|0_{\rm 3s}\rangle$ ($\frac{1}{\sqrt{2}}(|xxx\rangle+|\bar{x}\bar{x}\bar{x}\rangle)$, denoted as dot 4) or $|1_{\rm 3s}\rangle$ ($\frac{1}{\sqrt{2}}(|xxx\rangle-|\bar{x}\bar{x}\bar{x}\rangle)$, denoted as dot 6) which suggests that there is no off-diagonal
elements in the braiding matrix, $U_\text{ex}$. In addition, the relative
geometric phase between $|0_{\rm 3s}\rangle$ and $|1_{\rm 3s}\rangle$ during
the exchanging can be directly measured. Indeed, the final
states are obtained by rotating the initial states counterclockwise
along the {\bf X} axis through an angle of $\pi/2$ when the initial
state is a superposition state of $|0_{\rm 3s}\rangle$ and $|1_{\rm 3s}\rangle$, i.e., the basis $|0_{\rm 3s}\rangle$ obtains an extra phase factor of $-i$ relative to $|1_{\rm 3s}\rangle$.
Thus, the braiding matrix can be determined
by the relative geometric phase giving $U_\text{ex}=\text{diag}(1,e^{-i\pi/2})$. The braiding
matrix determined here agrees well with the theoretical result~\cite{Alicea2011}.
This implies that any input state $\alpha|xxx\rangle+\beta|\bar{x}\bar{x}\bar{x}\rangle$,
with the complex coefficients of $\alpha$ and $\beta$ ($|\alpha|^2+|\beta|^2=1$),
would be transformed to $\frac{1}{\sqrt{2}}(\alpha-i\beta)|xxx\rangle+\frac{1}{\sqrt{2}}(\beta-i\alpha)|\bar{x}\bar{x}\bar{x}\rangle$, i.e., the state $\frac{1}{\sqrt{2}}(\alpha+\beta)|0_{\rm 3s}\rangle+\frac{1}{\sqrt{2}}(\alpha-\beta)|1_{\rm 3s}\rangle)$ would be changed to $\frac{1}{\sqrt{2}}(\alpha+\beta)|0_{\rm 3s}\rangle+i\frac{1}{\sqrt{2}}(\alpha-\beta)|1_{\rm 3s}\rangle)$ in the basis of $|0_{\rm 3s}\rangle$ and $|1_{\rm 3s}\rangle$.

The real and imaginary parts of the experimentally determined
operator of the exchange process (the computation basis are represented as $|xxx\rangle$ and $|\bar{x}\bar{x}\bar{x}\rangle$), as determined via the quantum process tomography (represented by Pauli operators $\{I,X(\sigma^{x}),Y(\sigma^{y}),Z(\sigma^{z})\}$)
~\cite{OBrien2004},
are presented in Figures~\ref{fig:finalstate}\textbf{c}
and \textbf{d}. The fidelity is calculated to be $94.13\pm0.04$\% (the errors are deduced from the Poissonian photon counting noise).

If the statistics between the MZMs were Abelian then the braiding matrix between these two degenerate states would have been the identity matrix without any relative phase between the degenerate basis states. That we obtain a relative phase factor is in agreement with the non-Abelian character of MZMs.

\begin{figure*}[tbph]
\begin{centering}
\includegraphics[width=1\columnwidth]{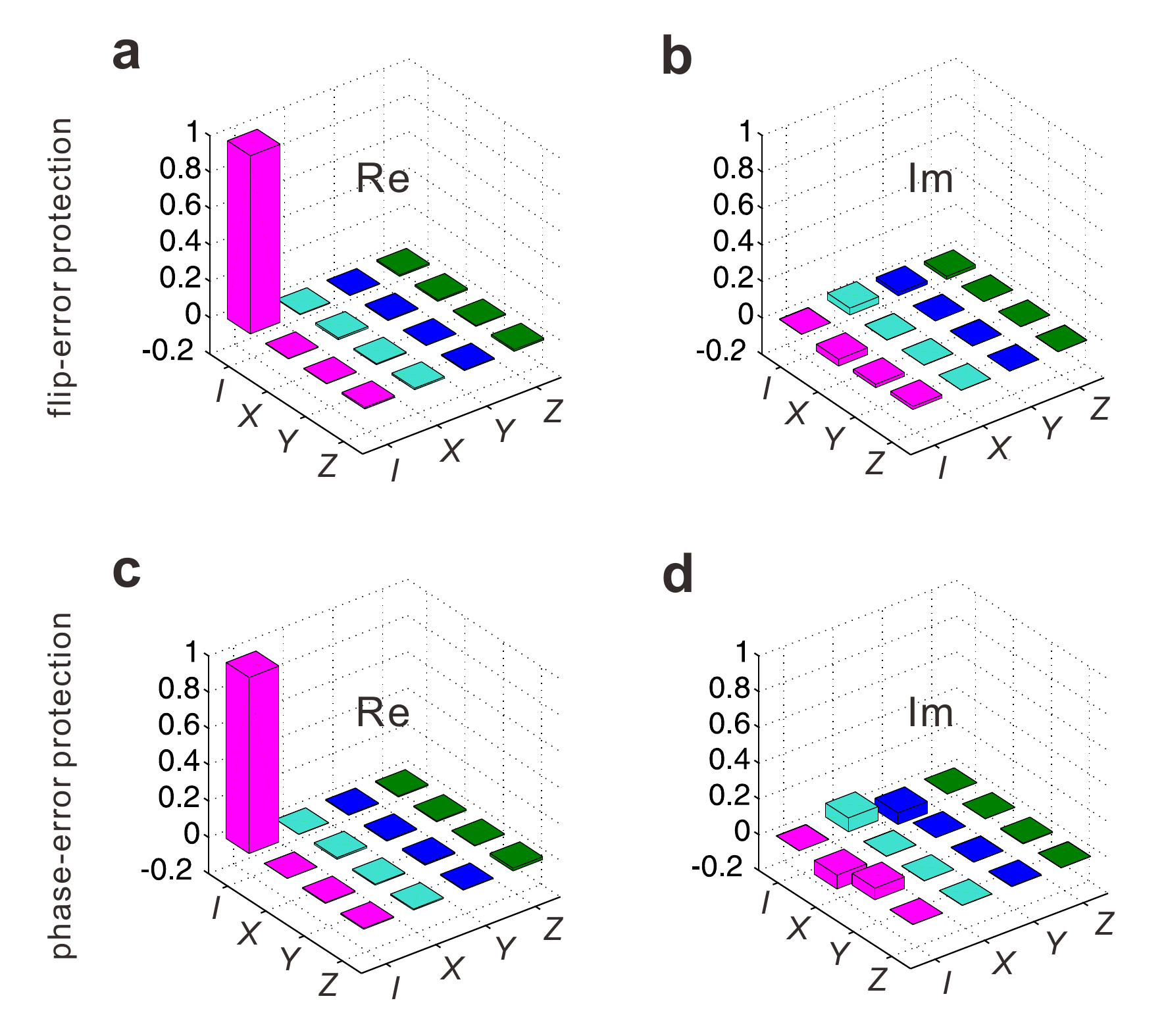}
\par\end{centering}

\protect\caption{Experimental results on simulating local noises immunity. \textbf{a.} Real (Re) and \textbf{b.} Imaginary (Im) parts of the flip-error protection operator, with a
fidelity of $97.91\pm0.03$\%. \textbf{c.} Real (Re) and \textbf{d.} Imaginary (Im) parts of the phase-error protection operator, with a
fidelity of $96.99\pm0.04$\%. The basis are expressed as $\{{I} ({\rm identity}), {X} (\sigma^{x}), {Y} (\sigma^{y}), {Z} (\sigma^{z})\}$}.
\label{fig:disturbancedata}
\end{figure*}

{\bf \noindent Experimental results on simulating local noises immunity.}
We further investigate local noises immunity of the information encoded in the ground space of the KCM. The output state after the dissipative evolution DE0, $\alpha|xxx\rangle+\beta|\bar{x}\bar{x}\bar{x}\rangle$, is treated as the initial state. After the operation of $\frac{1}{4}(i\sigma_{1}^{y}\sigma_{2}^{x}+\sigma_{1}^{y}\sigma_{2}^{y}+\sigma_{1}^{x}\sigma_{2}^{x}-i\sigma_{1}^{x}\sigma_{2}^{y})$ (flip-error operator), the two output modes would become eight. The final state is the same as the initial one by projecting the state back to the ground-space of $H_{0}$ (i.e., the operation of DE0 and we omit the unobservable overall phase). Similarly, the initial two output modes ($\alpha|xxx\rangle+\beta|\bar{x}\bar{x}\bar{x}\rangle$) become four modes after the local phase noise operation of $\frac{1}{2}(\sigma_{1}^{z}+1)$. The final state is projected back to the initial one with the ITE operation of $e^{-H_{0}t}$ (see section IF in SI for more details). In our experiment, the system is subjected to a projection back onto the initial state after applying the errors. By doing so, the success probability is small since the preserved state is just a fraction of the total state. However, it serves well for establishing the technology and methodology for the in-principle demonstration of the anyonic physics. Figure~\ref{fig:disturbancedata}\textbf{a}
and \textbf{b} show the real and imaginary parts of the flip-error protection operator with a fidelity of $97.91\pm0.03$\%.  Figure~\ref{fig:disturbancedata}\textbf{c}
and \textbf{d} show the real and imaginary parts of the phase-error protection operator with a fidelity of $96.99\pm0.04$\%. The errors of the fidelity values are deduced from the Poissonian
photon counting noise. This fact
reveals the protection from the local flip error and phase error in the
KCM. The total operation behaves as an identity, thus demonstrating immunity against noise.

Generally, the fidelity of protection operators are also affected by the noises without $\rm Z_2$ symmetry and the evolution time in the ITE. Due to the fact that the environment can be well controlled in our optical simulator, the setup established here can also be used to investigate the effect of the noise without $\rm Z_{2}$ symmetry in the ground space of the TFIM. Our simulator offers the possibility to study the effect of other types of noise, which can be presented in the real system supporting MZMs (superconductor-based system). A central advantage of the optical simulation is that all the implementations can be realised with high precision.

\section*{Discussion}

In this paper we experimentally investigated the physics of MZMs emerging in the KCM by simulating it with an equivalent photonic system. The photonic quantum simulator, which employs a set of projective measurements, represents a general and powerful approach to directly study quantum evolutions. We were able to transport two MZMs at the endpoints of a chain and measure the geometric phases resulting from their exchange. Moreover, we experimentally demonstrated that the degenerate subspace of the MZMs is resilient against local perturbations.

Our simulation is based on a non-local JW transformation that maps the fermionic KCM to an equivalent spin model. This transformation, while it changes the states of the system, it does not alter the associated unitary evolutions, thus allowing us to measure directly the braiding matrix of MZMs. The appeal of our photonic quantum simulator is based on the fact that it allows for the transport of MZMs. It offers the exciting future possibility to investigate the braiding character of muti-MZM systems and demonstrate their non-Abelian statistics.

Applications towards topological quantum computation may be possible by employing the versatile and well-controlled optical quantum simulator presented here. MZMs and their braiding cannot be used to perform universal quantum computations~\cite{Kitaev2008}. However, one can envision implementing specifically tailored quantum algorithms, such as the Deutch-Josza algorithm~\cite{Kraus2013} where the information will be topologically protected during the whole process.

Due to the fact that we encode the multi-site spin-1/2 system in the spatial modes of a single photon, the scalability of this method is limited. Nevertheless, it serves well for demonstrating the fundamental statistical properties and robustness of MZMs with such a small system size. Our main focus is on the physics and establishing the technology and the methodology.
The gained knowhow (investigating the braiding of anyonic quasiparticles through the imaginary-time evolution) can be picked up by other technologies that offer scalability, such as the trapped ions~\cite{Barreiro2011} or Josephson junctions~\cite{Devoret2013} system.
\section*{Methods}
{\bf \noindent The imaginary-time evolution operator on states:}
For a given Hamiltonian $H$ with a complete set of eigenstates $|e_{k}\rangle$
and the corresponding eigenvalues $E_{k}$, any arbitrary pure state
$|\phi\rangle$ can be expressed as
\begin{equation}
|\phi\rangle=\sum_{k}q_{k}|e_{k}\rangle,
\end{equation}
with $q_{k}$ representing the corresponding complex amplitude. Here,
we focus on pure states, but the argument is also valid for mixed
states. The corresponding imaginary-time evolution (ITE) operator
($U$)~\cite{Vidal2007} on the state becomes
\begin{equation}
\begin{split}
U|\phi\rangle & =\exp(-H*t)\sum_{k}q_{k}|e_{k}\rangle \quad \\
& =\sum_{k}q_{k}\exp(-tE_{k})|e_{k}\rangle.
\end{split}\label{imaginaryevolution}
\end{equation}
The evolution time $t$ is chosen to be 5 in our analysis, which is long enough to
drive the input state to the ground state of $H$.
After the ITE, the amplitude $q_{k}$ is changed to be $q_{k}\exp(-5E_{k})$.
We can see that the decay of the amplitude is strongly (exponentially)
dependent on the energy: the higher the energy, the faster the decay of the
amplitude. Therefore, only the ground states (with lowest energy)
survives during the evolution. Furthermore, due to the fact
that the Hamiltonian $H_{0}$, $H_{1}$ and $H_{2}$ can be divided
into two commuting parts, we can separate each ITE operator to two ITE
operators whose eigenvectors can be directly obtained (see Equation~(\ref{eq:commutation})).

{\bf \noindent Experimental process in probing braiding characteristics:}
The experimental setup for probing braiding characteristics is shown
in Figure~\ref{fig:setup_nonabelian}. The process follows the logical
diagram presented in the pane enclosed in the black solid line, denoted
by Logi. The setup used to prepare the input state is illustrated
in the pane labeled Pre. For simplicity, at the beginning of the evolution,
all three qubits are expressed in the basis of $\sigma^{x}$, in which
the ground states of $H_{0}$ are encoded in the two spatial modes
$|xxx\rangle$ and $|\bar{x}\bar{x}\bar{x}\rangle$. The dissipative evolution of $H_{0}$
with arbitrary input states is illustrated in the pane labeled DE0,
the output state of which is treated as the initial state. The basis
rotation BR1 is employed, and the subsequent dissipative evolution
(DE1) should drive the state to approximate the ground state of
$H_{1}$.

After BR2 and DE2, the two input modes are expanded to four output
modes, which are then sent to BR3 to implement the dissipative evolution
DE0. In our experiment, we perform two types of measurements, i.e.,
a two-mode measurement (TM) and a four-mode measurement (FM). The
initial states, which contain spatial information, are transformed
into states that contain polarization information, on which standard
quantum-state tomography of polarization states can be performed.
The TM corresponds to the tomography of the polarization states of
a single qubit, for which four types of polarization measurements
are required: horizontal polarization ($H$), vertical polarization
($V$), right-hand circular polarization ($R=1/\sqrt{2}(H-iV)$) and
diagonal polarization ($D=1/\sqrt{2}(H+V)$). The polarization-analysis
setup is constructed using a quarter-wave plate (QWP), a HWP and a
PBS. The photon is then detected using a single-photon detector (SPD).
Similarly, the FM corresponds to the measurement of two-qubit polarization
states, and 16 measurements are required. In the TM and FM, beam splitters
(BSs) are used to send photons to different measurement instruments.
The preparation of the initial states for the demonstration of noise
immunity is the same as is shown in Pre and DE0. The disturbance operation
is then implemented. After a second DE0, a TM is used to reconstruct
the final states (see sections IIA and IIB in SI for more details).

{\bf \noindent Data availability:}
The data that support the findings of this study are available from the authors on request.

\section*{Acknowledgments}

This work was supported by the National Plan on Key Basic Research and Development (Grants No. 2016YFA0302700), the Strategic Priority Research Program (B) of the Chinese Academy of Sciences (Grant No. XDB01030300), National Natural Science Foundation of China (Grant Nos. 11274297, 11105135, 11474267, 11004185, 60921091, 11274289, 10874170, 61322506, 11325419, 61327901), the Fundamental
Research Funds for the Central Universities (Grant No. WK2470000020), Youth Innovation Promotion Association and Excellent Young Scientist Program CAS, Program
for New Century Excellent Talents in University (NCET-12-0508) and Science
foundation for the excellent PHD thesis (Grant No. 201218).

\section*{Author Contributions}

Y.-J.H. constructs the theoretical scheme. J.-S.X. and C.-F.L. design the experiment. K.S. and J.-S.X. perform the experiment. J.K.P. contributes to the theoretical analysis. J.-S.X., Y.-J.H and J.K.P. wrote the manuscript. Y.-J. H. supervised the theoretical part of the project. C.-F.L. and G.-C.G. supervised the project. All authors read the paper and discussed the results.

\section*{Additional information}

The authors declare no competing financial interests.

\clearpage{}

\end{document}